%% file: main.tex
\documentclass[conference]{IEEEtran}
\IEEEoverridecommandlockouts
\usepackage{fontspec}
\usepackage{amsmath,amssymb}
\usepackage{graphicx}
\usepackage{booktabs}
\usepackage{array}
\usepackage{url}
\usepackage{xcolor}
\usepackage[hidelinks]{hyperref}
\newcommand{\pol}[1]{\textsf{#1}}

\input{posting-notice.tex}
\renewcommand{\baselinestretch}{0.97}

\begin{document}

\title{Where the Evidence Lives: Measuring Privacy and\\ Diagnosability of Exported AI-Agent Traces\thanks{Author manuscript accepted to S2AI 2026, co-located with IEEE BigData 2026. \ieeepostnotice}}

\author{\IEEEauthorblockN{Chang Xu}
\IEEEauthorblockA{\textit{Duke University}\\
chang.xu@alumni.duke.edu}
\and
\IEEEauthorblockN{Hao Wang}
\IEEEauthorblockA{\textit{Stanford University}\\
wanghao@alumni.stanford.edu}
\and
\IEEEauthorblockN{Gedi Zhou}
\IEEEauthorblockA{\textit{Independent Researcher}\\
gzhou4@alumni.jh.edu}
\and
\IEEEauthorblockN{Junyuan Tan}
\IEEEauthorblockA{\textit{University of California, Berkeley}\\
junyuan\_tan@berkeley.edu}}

\maketitle

\begin{abstract}
Execution traces help diagnose AI-agent failures, but sharing them can expose customer identities. For example, detecting that an agent handled the wrong customer’s records requires knowing whether the records belong to the requested customer, but not necessarily knowing who that customer is. Our executable benchmark measures what a trace export reveals and whether a proposed repair restores the correct output in synthetic multi-tenant workflows. Conventional redaction can remove evidence needed for repair while leaving encoded identifiers recoverable. Two exports preserve ownership comparisons by reporting the match as a comparison fact or by using pseudonyms consistent within each execution. Both preserve the original diagnostic program’s repairs. The tested identifier matchers identify no requested customer from either export. An independently written program produces correct outputs on 39 of 42 faulty cases with comparison facts and on all 42 with execution-scoped pseudonyms. Row counts still expose identities. With prior knowledge of customers’ row counts, an attacker uniquely identifies the requested customer in 97 of 240 executions under each of the 15 export policies. Removing selected counters from the comparison-fact export stops this attack and preserves the original program’s repairs. The benchmark lets engineers test what their diagnostic program can repair and what an attacker can still infer from the same export.
\end{abstract}


\section{Introduction}
When an AI agent on shared data infrastructure returns the wrong customer's records, an engineer must decide how much of its execution trace to share for diagnosis. The recipient needs to compare the request with the records to diagnose the ownership error. Sharing both enables that comparison but exposes identities. Deleting customer strings can remove the comparison and still leave an encoded identifier in a notification or tool argument.

Engineers need to know what the recipient can fix and what an attacker can infer. Even a correct diagnosis can lead to an edit that leaves the output wrong. We measure disclosure and check the output after running the proposed repair.

Existing work studies agent leakage and sanitization costs~\cite{elyagoubi2026agentleak}, retained audit information~\cite{xu2026redact} and diagnostic telemetry~\cite{zhu2026telemetrysuffbench,balusu2026agenttelemetry}. OpenTelemetry warns of personal data in captured inputs and outputs~\cite{otel2026genai}. Personally identifiable information detectors~\cite{presidio,scrubadub}, schema redaction and collector processors~\cite{otel2026sensitive} provide removal mechanisms. Our contribution is an executable evaluation that pairs these mechanisms' remaining disclosure with diagnosis and reference-scored outputs on the same executions.

A trusted exporter prepares the trace for sharing. It can report whether records belong to the requested customer, a result we call a comparison fact, or use pseudonyms that let the recipient make the comparison. Both preserve the repairs of the original rule-based program. An independently written program succeeds on 42/42 faulty executions with execution-scoped pseudonyms and 39/42 with comparison facts. The difference shows why exporters should test repair with the program that will use the export. In the main study, execution-scoped pseudonyms also make traces easier to link in one workflow.

\section{One Execution, from Export to Outcome}\label{sec:example}
In this scored example, a workflow reads a CSV file and writes the selected records as JSON lines. The filter selects customer B, so the output contains 16 of B's records. The expected output contains 64 records from the requested customer A. The letters A and B stand in for customer identities in this example.

Here the exporter prepares five versions of the trace. A diagnostic program reads each export and proposes an edit if it finds enough evidence. A trusted executor applies the edit and reruns the workflow in a copy of the original workspace, which contains the workflow's data and files.

Table~\ref{tab:example} follows the same case through five exports. Raw makes the request-to-record mismatch visible. Regex replaces both identifiers with redaction placeholders, which the diagnostic program treats as unknown rather than proof of equality. Schema removes the requested identity and row-owner fields. Comparison facts say that none of the rows belong to the requested customer and that all rows have the same owner. Execution-scoped pseudonyms give the request and parsed rows different tokens. Facts delegate the comparison to the exporter, while pseudonyms let the diagnostic program derive it.

\begin{table}[t]
\caption{One existing wrong-customer case. A/B are display aliases. Output and normalized matching are separate scored endpoints.}\label{tab:example}
\centering\footnotesize\setlength{\tabcolsep}{3pt}
\begin{tabular}{@{}>{\raggedright\arraybackslash}p{1.5cm}>{\raggedright\arraybackslash}p{2cm}>{\raggedright\arraybackslash}p{1.8cm}>{\raggedright\arraybackslash}p{2.1cm}@{}}
\toprule
Export & Ownership evidence & Action and output & Target disclosure \\
\midrule
Raw & Request A, rows B & Edit / 64 A rows & Exposed, tied top score \\
Regex & Comparison unknown & No-op / 16 B rows & Exposed, sole top score \\
Schema & Comparison unknown & No-op / 16 B rows & Exposed, tied top score \\
Facts & No row owner matches request & Edit / 64 A rows & No direct exposure or unique match \\
Execution-scoped pseudonyms & Request token differs from row token & Edit / 64 A rows & No direct exposure or unique match \\
\bottomrule
\end{tabular}
\end{table}

For the raw trace, comparison facts and execution-scoped pseudonyms, the rule-based program sets the parser's \texttt{owner\_param} to \texttt{requested}. The executor resolves that option using the original request, so the program can choose the correct edit without seeing A's identifier. The three exports yield the same edit, which is applied once. The corrected output contains 64 records from A and passes the record, ownership, checksum and source-integrity checks. With regex and schema redaction, the program makes no edit and the output still contains B's 16 records.

The privacy score tells a different story. Regex and schema both leave A's encoded e-mail in a notification. The matcher decodes it and compares it with the customer catalog. A ties with two other customers under raw and with B under schema. Regex removes the competing matches, so A is uniquely identified. Neither comparison facts nor pseudonyms yield a direct target match or unique identification by this matcher. Section~\ref{sec:results} tests repair and disclosure across the cohort.

\section{Benchmark and Evaluation}\label{sec:bench}
\subsection{Roles, Outputs and Evidence}
The diagnostic program sees the export and a public list of allowed operations. It may choose at most two configuration operations at one stage. With an edit, the executor reruns the entire workflow on the copy. Without an edit, the original output is scored. The benchmark seals the decision before scoring the result against a private reference output.

\emph{Task success} means that the actual output passes the reference scorer. \emph{Recovery} means that an initially failing workspace becomes correct. These denominators differ because some injected faults leave the original output correct. We also distinguish structural evidence, such as scan and write settings, value evidence, such as amount changes, and ownership evidence, such as whether records belong to the requested customer.

Three author-constructed workloads use synthetic identities and faults with real file, SQLite and CSV operations. The \emph{file workflow} parses CSV and delivers JSONL with a checksummed manifest. The \emph{reconciliation workflow} imports rows, reconciles customer totals and exports CSV. The \emph{aggregation workflow} reads partitioned transactions, joins a dimension table, aggregates decimal amounts and writes a report. Scoring checks canonical records and ownership, reconciliation source-table row counts, and file/aggregation source or input hashes.

Each workflow has seven fault families, with two variants per family: adapter settings, stale bindings, incomplete scans, wrong write mode, changed values, wrong customer and cross-customer contamination. The public edit schema permits a correction for each variant. The oracle applies that correction, while the no-op control preserves the original output.

A \emph{cell} consists of a workflow, seed and input instance. Each cell has a clean case and seven faults. For each workflow and seed, the first input uses variant 1 for all seven families and the second uses variant 2. The eight cases share a target, workspace and 32-candidate catalog. Only the target and two neighbors own rows. The other candidates are zero-volume decoys. Notifications represent identities as URL-query e-mail, base64 JSON headers, split e-mail/identifier fields or at/dot-obfuscated text. Diagnostic programs and attackers receive no private target labels or attack-time secrets. The count attacker additionally knows candidate volumes.

\subsection{Exports and Diagnostic Programs}
Table~\ref{tab:key} groups the 15 policies by what the diagnostic program receives and maps descriptive names to the frozen artifact identifiers.

\begin{table}[t]
\caption{Export families and terminology. Names in parentheses map to unchanged artifact identifiers.}\label{tab:key}
\centering\footnotesize\setlength{\tabcolsep}{3pt}
\begin{tabular}{@{}>{\raggedright\arraybackslash}p{2.35cm}>{\raggedright\arraybackslash}p{5.6cm}@{}}
\toprule
Family or role & Meaning and artifact mapping \\
\midrule
Raw and scrubbed content & Raw, regex and schema exports, plus default Scrubadub and Presidio. \\
Content reduction & Typed projection or metadata-only export, with full metadata or two metadata ablations. \\
Comparison facts & Request/record and stage-pair comparisons (\pol{rel}), optionally with run-scoped or execution-scoped owner tokens (\pol{rel-token}, \pol{rel-salted}). \\
Identifier substitution & Literal dictionary tokens (\pol{dict-tok}) or encoding-aware pseudonyms scoped to a run or an execution (\pol{canon-global}, \pol{canon-exec}). \\
Rule-based program & Ownership, metadata and value rules (C1). Metadata-only rules are C0. \\
Constraint-based program & Independently written constraint-based task decisions (C4). \\
Exploratory model classifier & Model-based fault classification (C3). \\
Disclosure tests & Basic/normalized matching (A0/A1), generic/identifier-aware linkage (A2/A2b), known-volume matching (A3). \\
\bottomrule
\end{tabular}
\end{table}

All full-metadata policies retain baseline measurements, statistics and relations. Measurements include counts, exceptions, scan arguments, digests and reference equality. Statistics describe key multiplicity and string-length buckets. Relations describe cardinality order and key coverage. Two metadata-only ablations remove relations, then statistics.

Raw retains requests, records, tool calls and byte snapshots. Regex applies five frozen patterns for e-mail, 32-hex keys, phone, SSN and API keys, including decoded base64 byte fields. Schema removes \texttt{owner\_id}, \texttt{owner\_email}, \texttt{billing\_id}, \texttt{note} and byte blobs. Typed projection retains stage summaries of counts, status, scan, write mode, reference equality and key multiplicity. The metadata-only export retains the measurements, statistics and relations described above. The post-hoc scrubber baselines use default Scrubadub~\cite{scrubadub} detectors and Presidio~\cite{presidio} recognizers.

Comparison facts add content without fault labels. At each stage they report whether owners match the request (all, some, none, unknown or not applicable), distinct-owner count, a Boolean predicate for whether all amounts are whole dollars, and negative-amount count. Stage-pair edges retain endpoints, report whether transaction counts agree, and classify amount sums as equal, negated, scaled by 100, less or greater. Unavailable amount comparisons are null. This added fact content omits raw sums and transaction counts but retains owner and negative-amount counts. Baseline metadata still retains counts, including the filtered row count used by the volume attacker. Optional owner tokens use a hash-based message authentication code with a split/run key or an execution-handle key that attackers do not receive.

The encoding-aware pseudonym policies transform the original content. They map complete owner identifiers and e-mails consistently across all four notification forms and decoded payloads, with an analysis-run key or an execution-handle key. Both retain baseline metadata, original snapshot counts/digests and transformed-payload metadata. In contrast, dictionary substitution does not consistently canonicalize notifications, and the fact-token variants add owner tokens to facts rather than retaining transformed records. Scrubadub, Presidio, dictionary substitution and both pseudonym scopes are the five post-hoc comparators, evaluated on the same executions.

The rule-based program first checks ownership, then metadata, then values. It derives comparisons from visible content or reads exported facts. Equal counts with unequal sums trigger sign, scale or rounding edits. For the file workflow's source-row sum, it needs the visible requested owner to select records. Missing evidence is unknown. It ignores distinct-owner and negative-amount counts. Because this diagnostic program and the facts share a family-selected vocabulary, agreement alone is an incomplete test of usefulness. The independent constraint-based program tests whether another decision procedure can still repair (Sec.~\ref{sec:independent}).

\subsection{Disclosure Endpoints}
Identifier matching ranks the 32 catalog candidates. Basic matching searches raw, lower-case, URL, URL-plus, base64 and hex forms plus length buckets. Normalized matching adds Unicode compatibility normalization, case, whitespace and at/dot handling. It credits full atoms, e-mail local parts and 12-character identifier fragments. \emph{Unique identification} requires the requested target alone to score highest. \emph{Direct exposure} requires a score of at least 10, reached by full identifier/e-mail matches regardless of ties. Fragment and shape matches rank candidates below that threshold. Thus exposed neighbors can reduce raw unique identification by creating ties. Target and neighbor exposure are scored separately. Uniform catalog guessing has reference probability $1/32$, or $1/3$ among the three row owners.

Linkage asks whether two traces share a customer and input cell. Generic linkage uses Jaccard similarity after removing near-universal strings in each same-policy batch. Identifier-aware linkage uses one frozen identifier-shape grammar on content and baseline metadata, including eligible original/transformed digests but excluding auxiliary-only strings and other token forms. A separate post-hoc dictionary-token extractor admits \texttt{tok:} strings. The area under the receiver operating characteristic curve (AUC) measures separation of positive same-cell and negative different-cell pairs within each workflow. Shared inputs make this customer/input linkage rather than isolated identity linkage.

Known-volume matching compares the number of rows selected by the customer filter, before any scan limit or offset, with each catalog customer's known row count. The export records this count as \texttt{rows\_in}. Its auxiliary information associates each candidate identity with a count, including the three row owners and zero-volume decoys. It succeeds only if the requested target is the sole count match. This endpoint tests a metadata channel beyond string matching.

\subsection{Cohorts and Protocol}\label{sec:protocol}
The main scripted study uses disjoint seeds for 576 development, 48 validation and 240 final executions, comprising 210 faults and 30 clean cases. Ten original policies produce 2,400 final views. A master secret derives identities, forms and tokens and is released after scoring. Diagnosis and matching attacks use public views without target labels. A post-hoc replay scores retained final inputs under all 15 policies and a selected-counter fact ablation, with oracle/no-op controls. Identical execution/plan pairs share scored outputs.

The independent diagnostic program uses a separate 48-case cohort. A primary 144-execution language-model cohort and a separate post-hoc 144-execution cohort test generated traces and interaction exposure. Counts describe finite cohorts, with shared cells limiting population inference. Artifact documentation preserves the full per-policy results.

\section{Repair and Remaining Disclosure}\label{sec:results}
\subsection{Why Redaction Can Lose Evidence and Still Disclose}
Most structural diagnosis survives in metadata, but ownership and some value diagnosis require content comparisons. The metadata-only rules make identical decisions across the eight original policies with full metadata (0/240 disagreements). They detect 115/120 structural faults, falling to 110/120 without relations and 75/120 with measurements alone. The measurements-only view detects no write-mode faults. The rule-based program adds five amount-scale detections under raw, regex, schema and facts. Thus even a predominantly structural group contains a value-dependent case.

The ownership comparison creates the clearest repair loss. The rule-based program detects all 60 wrong-customer/contamination faults under raw and facts, and none under regex, schema or metadata-only exports. Regex and schema detect 20/30 value faults because removing the requested owner prevents the file workflow's source-row comparison. Facts retain the exporter-computed result. Comparison facts and both pseudonym scopes achieve 210/210 fault-task successes and 190/190 recoveries, compared with regex/schema's 140/210 and 120/190 (Table~\ref{tab:main}).

\begin{table}[t]
\caption{Eight selected exports on the main cohort. Fault success is out of 210 cases, and recovery is out of 190 initial failures. Normalized identification and target exposure are out of 240 cases. All rows retain full metadata.}\label{tab:main}
\centering\footnotesize\setlength{\tabcolsep}{3.3pt}
\begin{tabular}{@{}lcccc@{}}
\toprule
Export & Success & Recovery & Unique ID & Exposure \\
\midrule
Raw & 210 & 190 & 65 & 240 \\
Regex & 140 & 120 & 225 & 163 \\
Schema & 140 & 120 & 221 & 226 \\
Scrubadub* & 210 & 190 & 153 & 240 \\
Presidio* & 184 & 164 & 202 & 235 \\
Metadata-only & 115 & 95 & 0 & 0 \\
Comparison facts & 210 & 190 & 0 & 0 \\
Execution-scoped pseudonyms* & 210 & 190 & 0 & 0 \\
\bottomrule
\end{tabular}\\[2pt]
\parbox{\columnwidth}{\scriptsize *Post-hoc, same cases. Raw and Scrubadub often expose neighbors, creating top-score ties. Zero matching is specific to the tested grammar. Every policy still permits 97/240 unique known-volume matches.}
\end{table}

Removing repair evidence does not ensure anonymity. Normalization increases regex unique identification from 49/240 under basic matching to 225/240, while schema reaches 221/240. Residues include filter arguments, owner stamps, transaction keys and unlabeled CSV. Raw uniquely identifies only 65/240 because of target/neighbor ties, despite exposing the target in every case. The metadata-only export, comparison facts and both encoding-aware pseudonym policies yield zero unique identification and direct exposure under both matching attacks. The pseudonym policies were written for the same declared identity grammar, so these zeros measure coverage of that grammar.

Additional scrubbers show the same tradeoff. Scrubadub preserves rule-based program success but exposes opaque account identifiers as often as raw. Default Presidio retains identifiers that enable both disclosure and ownership checks. It detects 192 faults, localizes 183 and succeeds on 184. Dictionary substitution leaves 161/240 normalized unique matches, and its separate token-aware linkage extractor reaches AUC 1.0000 in every workflow.

Output scoring distinguishes diagnosis from repair. All policies pass 30/30 clean cases, including Presidio despite two false alarms. Oracle and no-op pass 210/210 and 20/210 faults. Measurements alone yield 75 detections but 80 successes because five undetected faulty executions retain correct outputs through no-ops. Recovery excludes these initially correct executions.

\subsection{Retention with an Independent Program}\label{sec:independent}
The original rule-based program succeeded with both comparison facts and pseudonyms. To see whether another program can use both exports, we test a separately written program that chooses edits using the export and public workflow rules. These rules give allowed settings and defaults, such as filtering for the requested customer. Any ticket contains only the retained request. The program enumerates allowed one-stage configurations, keeps plans addressing a listed candidate cause for every observed constraint, and ranks them by fewer operations then fewer unrelated changes. It acts only on a unique best plan. Missing or ambiguous support causes abstention, without workspace probing.

On raw development traces, the program succeeds on 21/21 faults and 3/3 clean cases. Its source and criteria were frozen before the 48-case test, which uses two seeds per workflow, first fault variants and clean controls. On raw test traces, the program must succeed on at least 36/42 faults, including 21/24 structural, 5/6 value and 10/12 ownership cases, plus 6/6 clean. To retain repair, both facts and execution-scoped pseudonyms must lose at most two fault successes relative to raw, meet the value, ownership and clean minima, and raise at most one clean false alarm. Abstentions and invalid plans count as no-ops.

\begin{table}[t]
\caption{Independent constraint-based program on its separate 48-case cohort. Fault success and recovery are distinct endpoints. Disclosure attacks were not run on this cohort.}\label{tab:independent}
\centering\footnotesize\setlength{\tabcolsep}{3.2pt}
\begin{tabular}{@{}lccc@{}}
\toprule
Export & Fault success /42 & Recovery /36 & Clean /6 \\
\midrule
Raw & 42 & 36 & 6 \\
Regex / schema & 28 & 22 & 6 \\
Comparison facts & 39 & 33 & 6 \\
Execution-scoped pseudonyms & 42 & 36 & 6 \\
\bottomrule
\end{tabular}
\end{table}

Raw and execution-scoped pseudonyms succeed on 42/42 faults and recover 36/36 initially failing cases. Facts succeed on 39/42 and recover 33/36 (Table~\ref{tab:independent}). The observed effect is three lost successes. Because the allowance is two, facts miss the retention criterion by one case. They retain 22/24 structural, 6/6 value and 11/12 ownership successes. Regex/schema achieve 24/24, 4/6 and 0/12 respectively. All five exports pass 6/6 clean cases without diagnostic false alarms or invalid plans. Oracle matches raw, while no-op passes six faults and recovers none.

The three fact losses are abstentions. Two file-workflow delimiter faults retain the parse exception but omit the delimiter setting needed to localize a constraint.

In a separate wrong-customer case, the program infers an irrelevant amount constraint from the facts. The facts report equal transaction counts but unequal sums between the source stage and parsed output. The diagnostic program therefore seeks to restore amount conservation as well as owner selection, although the wrong filter selected different customers' transactions. Its recorded constraints, public settings and ranking rules yield two equally ranked plans. Both set \texttt{owner\_param} to \texttt{requested}. One also disables rounding and the other resets the sign setting. These plans are statically reconstructed, not observed causes or executed repairs. Their tie causes abstention.

The raw export and the export with execution-scoped pseudonyms retain transaction identifiers. They show that the requested customer's source records and parsed records are different sets despite equal sizes. This diagnostic program therefore skips the amount comparison and chooses the unique one-operation owner correction. The summarized view retains the ownership mismatch, but the extra amount constraint prevents a unique decision under these rules. This interaction does not establish that comparison facts are intrinsically insufficient.

Raw and execution-scoped pseudonyms agree on every outcome, gaining three successes over facts without a reverse loss. Facts still outperform either regex or schema on 13 cases and lose on two, a net gain of 11. The rule and constraint-based programs use different cohorts, so their aggregate scores do not isolate the effect of changing the program alone.

\subsection{Linkage and Counts after Identifier Matching}
The main cohort shows linkage even when identifier matching fails. Execution-scoped pseudonyms and facts have equal identifier-aware AUC in the file and reconciliation workflows (0.8750 and 0.5000). In aggregation, pseudonyms reach 0.9633 versus facts' 0.6585. Run-scoped pseudonyms reach 1.0000 in all three. These feature sets include metadata digests. We did not isolate their contribution or test disclosure on the independent program's cohort.

To see how much linkage depends on the attacker's choice of features, we also test generic similarity on comparison facts with run-scoped owner tokens. Generic linkage gives AUCs of 0.7797, 0.7898 and 0.7698 in the file, reconciliation and aggregation workflows, below the 0.95 minimum predicted in our protocol. On the same exports, the identifier-aware attack reaches 1.0000, 1.0000 and 0.9971. Each workflow has 80 views from 10 shared cells, giving 280 positive and 2,880 negative dependent pairs.

Known-volume matching uniquely identifies 97/240 requested targets under every one of the 15 policies. A selected-counter fact ablation removes seven fields: measurement row/byte counts and statistic row/key/distinct-key counts. It retains baseline relations, distinct-owner count and negative-amount count. The count attack then has no match in 240/240 views, while rule-based program detection and fault-task success remain 210/210 and recovery remains 190/190. Identifier-aware linkage and outcomes for the independent program were not measured for this ablation. The result is specific to the stated volume rule.

\section{A Secondary Language-Model Check}\label{sec:agent}
The language-model study asks whether ownership diagnosis and identifier matching show the same contrast on model-generated traces. Qwen2.5-3B-Instruct (4-bit, CPU, greedy) follows a ticket/runbook and calls native stages. The environment injects faults beneath its arguments. There are 96 development and 144 primary held-out executions. Calls are valid in 912/967 steps (94\%), and all episodes complete the harness. Exports are evaluated by the same rule-based program and normalized matcher. A separate exploratory model classifier predicts seven fault families plus clean from at most 768 tokens. This study does not test model-based repair.

The written protocol specifies at least 90\% raw/fact ownership detection versus at most 10\% for regex/schema/metadata-only, at least 50\% normalized identification for regex/schema versus zero for metadata-only and fact variants, and at least 90\% fact detection within each evidence group. These criteria are met. Facts detect 70/72 structural, 17/18 value and 36/36 ownership faults, with zero normalized unique identification. Regex/schema detect 0/36 ownership faults but uniquely identify 113/144 and 137/144 targets. One eight-execution cell filters by e-mail instead of account ID and returns no rows. It remains in the primary analysis.

Interaction content supplies a separate exposure channel. Restricting normalized matching to step index, explanation, tool, arguments and response, and clearing other content and metadata, still exposes the target in 127/144 schema-redacted primary executions and 130/144 separate post-hoc executions. Among nine primary episodes whose nonempty typed filter rows belong solely to a neighbor, four accepted simulated \texttt{notify} calls have a neighbor-matching e-mail and two match both account and e-mail. The separate new-seed cohort repeats these 4/9 and 2/9 counts without an execution-changing agent error. These retained simulated fields do not establish external delivery.

Facts show no demonstrated improvement in exploratory model diagnosis. With invalid predictions counted incorrect, facts yield 30/144 correct predictions versus metadata-only's 26/144. There are six fact-only wins and two wins with metadata alone (two-sided paired exact McNemar $p=0.289$). Excluding the eight filter deviations removes both wins with metadata alone and changes the counts to 29/136 versus 23/136. That selected sensitivity does not replace the primary result. Shared cells make the episode-paired test nominal.

\section{Using the Benchmark and Interpreting Its Scope}
To compare exports on new inputs, a researcher passes each view and the public edit schema to a diagnostic program and seals its decision before private replay and scoring. A two-row file example shows this interface with raw, execution-scoped pseudonym and metadata-only exports. The example and aggregate results for all 15 policies are in \href{https://github.com/AnthonyXu109/s2ai-2026-artifact/releases/tag/v1.0.0}{release v1.0.0} at \url{https://github.com/AnthonyXu109/s2ai-2026-artifact}. Full study materials are available on request.

Exporters should audit encoded identifiers, counts and linkage separately, then check repaired outputs with their intended program and an independent implementation. Testing alternative count summaries and execution-keyed digests would extend these results.

The evidence comes from three synthetic workflows, seven injected fault families and bounded edits. The public defaults narrow the search for repairs. The rule-based program and facts share a selected vocabulary, while the independent diagnostic program tests only first fault variants on a different cohort. Pseudonyms cover the declared grammar searched by the matching attacks. Stronger auxiliary-data or model-based re-identification is outside these tests. The language-model evidence uses one small runbook-guided model and environment faults. Main and model studies retain protocols, manifests and deviations, but their prospective protocol timing could not be independently established. The independent diagnostic program's source and criteria were frozen before its own final cases.

\section{Related Work}
\emph{Trace anonymization.} Packet privacy~\cite{pang2006devil}, Crypto-PAn's consistent keyed IP mappings~\cite{xu2002cryptopan}, log anonymization~\cite{aghili2025logs} and unconventional identifier forms~\cite{zafar2026mindgap} concern trace publication risks. Secret Sharer tests memorization~\cite{carlini2019secretsharer}, while trace-canary audits test evidence of weight memorization~\cite{li2026canaries}. Our identification endpoint concerns the requested customer in one execution and catalog.

\emph{Agent privacy and auditability.} LOPSIDED pseudonymizes prompts and reconstructs responses~\cite{serenari2025lopsided}. We evaluate post-execution exports. AgentLeak tests internal leakage and sanitization costs~\cite{elyagoubi2026agentleak}. RedAct protects procedural knowledge and evaluates audit fields and human-rated usability~\cite{xu2026redact}. Source-aware redaction~\cite{zhou2023sourceaware}, identifier detectors~\cite{presidio,scrubadub} and OpenTelemetry~\cite{otel2026sensitive} supply mechanisms. AgentTelemetry counts telemetry-guided, agent-proposed, guard-checked patch interventions in a SWE-bench Lite case study~\cite{balusu2026agenttelemetry}. We pair replayed outputs with disclosure.

\emph{Failure and task evaluation.} AgentChaosBench tests runtime faults and sanitized-telemetry localization~\cite{zhang2026agentchaos}. TraceElephant supplies 220 filtered, manually inspected public-benchmark failure traces without our requested-customer labels~\cite{chen2026traceelephant}. TelemetrySuffBench tests paired-view failure-origin sufficiency and nonunique origins~\cite{zhu2026telemetrysuffbench}. We join diagnosis and post-decision output scoring with disclosure/linkage. AgentDojo~\cite{debenedetti2024agentdojo} and computer-use probes~\cite{li2026auc0998} test injection security. $\tau^2$-Bench tests task utility~\cite{barres2026tau2}. Our edits are bounded configurations.

\section{Conclusion}
An export can remove identifiers, lose repair evidence and still disclose the requested customer. Comparison facts and both pseudonym scopes preserve the rule-based program's recoveries. The independent program loses three repairs with facts and retains them with execution-scoped pseudonyms. On the main cohort, those pseudonyms permit stronger linkage in the aggregation workflow. Dropping selected counters from the comparison-fact export stops the tested row-count attack and preserves rule-based repairs. The benchmark lets exporters test repair and remaining disclosure on the same exports.

\section*{Acknowledgment}
Generative AI tools, including Claude, Muse and Codex, assisted with code development, experimental tooling, and drafting and revising Sections I--VIII. The authors are responsible for the article's content.

\bibliographystyle{IEEEtran}
\bibliography{refs-codex-v32}
\end{document}

%% file: posting-notice.tex
\newcommand{\ieeepostnotice}{\copyright\ 2026 IEEE. Personal use of this material is permitted. Permission from IEEE must be obtained for all other uses, in any current or future media, including reprinting/republishing this material for advertising or promotional purposes, creating new collective works, for resale or redistribution to servers or lists, or reuse of any copyrighted component of this work in other works.}